\documentclass{jfm}
\usepackage{graphicx}
\usepackage{amsmath}
\usepackage{overpic}
\usepackage{float}
\usepackage{epstopdf,epsfig}
\usepackage{color,contour}
\usepackage{amssymb}
\usepackage[dvipsnames]{xcolor}
\usepackage{graphicx}
\usepackage{tabularx}
\usepackage[dvipsnames]{xcolor}
\usepackage{booktabs}
\usepackage{ulem}
\usepackage{bm}
\usepackage[colorlinks,citecolor = blue, linkcolor=red,hyperindex,CJKbookmarks]{hyperref}

\title[Rotation of anisotropic particles in Rayleigh-B\'enard turbulence]{Rotation of anisotropic particles in Rayleigh-B\'enard turbulence}

\author[L. Jiang and others]%
{Linfeng Jiang$^1$
, Enrico Calzavarini$^{2,\thanks{Email address for correspondence: enrico.calzavarini@polytech-lille.fr}}$
, Chao Sun$^{1,3,\thanks{Email address for correspondence: chaosun@tsinghua.edu.cn}}$ \\}
\affiliation{ $^1$Center for Combustion Energy, Key laboratory for Thermal Science and Power Engineering of Ministry of Education, International Joint Laboratory on Low Carbon Clean Energy Innovation, Department of Energy and Power Engineering, Tsinghua University, Beijing 100084, China \\[\affilskip]	
 $^2$Univ.\ Lille, Unit\'e de M\'ecanique de Lille - J. Boussinesq - UML - ULR 7512, F-59000 Lille, France \\ [\affilskip]
 $^3$Department of Engineering Mechanics, School of Aerospace Engineering, Tsinghua University, Beijing 100084, China 
}

\pubyear{2010}
\volume{650}
\pagerange{119--126}
\date{?; revised ?; accepted ?. - To be entered by editorial office}
\begin{document}

\maketitle

\vspace{-2 mm}
\begin{abstract}
Inertialess anisotropic particles in a Rayleigh-B\'enard turbulent flow show maximal tumbling rates for weakly oblate shapes, in contrast with the universal behaviour observed in developed turbulence where the mean tumbling rate monotonically decreases with the particle aspect ratio.
This is due to the concurrent effect of turbulent fluctuations and of a mean shear flow whose intensity, we show,  is determined by the kinetic boundary layers. In Rayleigh-B\'enard turbulence  prolate particles align preferentially with the fluid velocity, while oblate ones orient with the temperature gradient. This analysis elucidates the link between particle angular dynamics and small-scale properties of convective turbulence and has implications for the wider class of sheared turbulent flows.
\end{abstract}
\vspace{-2 mm}


%----------------Introduction----------------
\section{Introduction}
Can we infer the small scale features of a turbulent flow from the angular dynamics of a particle advected by it? This question has represented a motivation and a formidable challenge for much of the research performed on Lagrangian studies of turbulence over the last decade \citep{VothARFM2017}.  The angular dynamics of a small material body advected by a flow indeed inherits properties of the spatial gradient of the carrying flow. However, even in the most simple instance, the one of an inertialess axisymmetric particle in a plane steady shear flow whose description is due to G. B. Jeffery \citep{Jeffery1922}, the connection is not straightforward owing to the non-linearity of the particle equation of motion. A particle performing a so called Jeffery orbit tumbles at variable speed and, as a result, on average preferentially aligns with certain flow directions. Notable progresses have been made in the context of statistically steady, homogeneous and isotropic turbulence (HIT), where a robust universal behaviour, \textit{i.e.}~independent of external forcing, has been highlighted \citep{Shin2005,PumirWilkinson2011,ParsaPRL2012, Gustavsson2014,ni_ouellette_voth_2014,ni_kramel_ouellette_voth_2015,Byron2015,CandelierPRL2016,PhysRevLett.121.044501}.
 
The study of rotation of anisotropic particles has known a great development in recent times also motivated by applications in environmental sciences, geophysics and industry. Anisotropic particles are encountered 
in the form of cloud ice crystals in atmospheric precipitation models \citep{heymsfield1977precipitation} or in the shape of planktonic organisms in ocean population dynamics research \citep{ardeshiri2017copepods} or as cellulose fibers in paper making industrial processes \citep{olson1998motion}, to cite only few application domains.
The advancement of knowledge has followed two main paths: i) refine the particle description including hydrodynamic forces due to inertia, finite-size and shape \citep{chevillard_meneveau_2013,ParsaPRL2014,KramelPRL2016,pujara_variano_2017,BounouaPRL2018}, or external forces such as gravity \citep{MarchioliPF2010,GustavssonPRL2017}; ii) explore different fluid flows, from laboratory-scale turbulence to geophysical settings. The second aspect is still largely uncharted, while many studies are available for the above mentioned HIT condition, only a handful of works cover the topics of bounded turbulence \citep{Mortensen2008,Marchioli2013,Zhao2015,Challabotla2015,Bakhuis2019}, or surface flows \citep{dibenedetto_ouellette_koseff_2018}. The fact that hitherto no study exists for thermally driven turbulence (so relevant for environmental and geophysical applications) provides a motivation for the present work.
Instances of thermally driven flows are, e.g., atmospheric winds and ocean currents, magma and earth mantle convection, and any molten flow in industry.

In this paper we show by means of experiments and numerical simulations that the tumbling rate of inertialess neutrally buoyant axisymmetric particles in a convective turbulence Rayleigh-B\'enard (RB) cell displays a peculiar trend as a function of the particle aspect ratio, which is different from the one in HIT. We demonstrate that this trend is due to the combined effect of turbulent fluctuations and a persistent shear flow determined by the kinetic boundary layers.
In the discussion we speculate that such behaviour will vanish for larger thermal forcing (higher Rayleigh number), finally recovering the universal behaviour observed in turbulence.

The paper is organized as follows. The experimental and numerical methods are described in detail in $\S \ref{sec:methods}$, including the governing equations of the flow and particles. We present the main results in $\S \ref{sec:results}$, focusing on the statistics of the tumbling rate \textcolor{black}{and preferential alignment of} anisotropic particles. Finally, in $\S \ref{sec:summary}$ we summarize our findings.
%sec_heat_transfer
%sec_flow_structure
%sec_energy_dissipation

\section{Methodology}\label{sec:methods}
\begin{figure}
	\begin{center}
		\includegraphics[width=0.8\columnwidth]{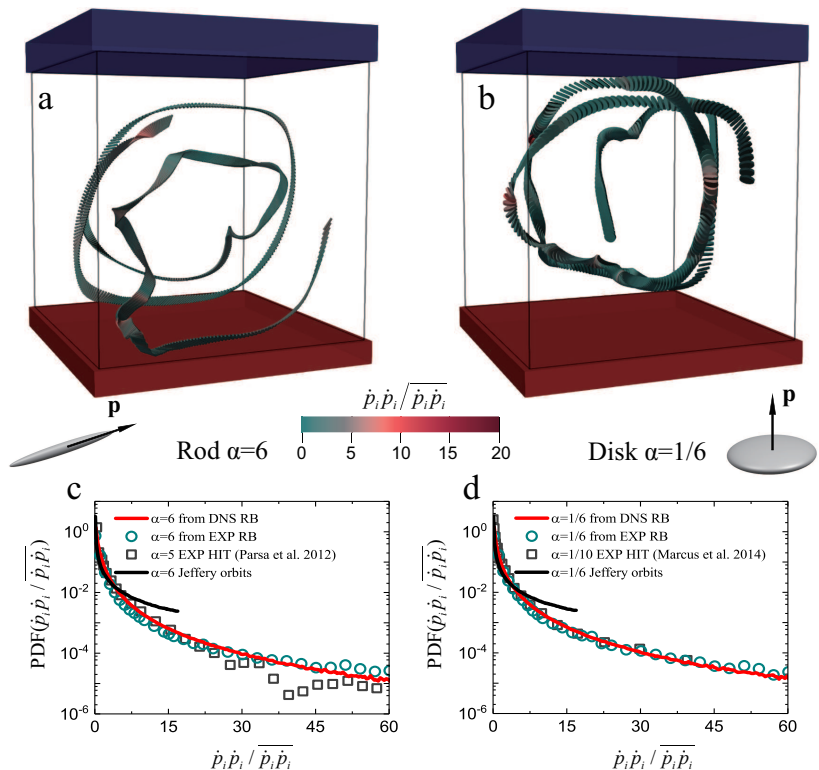}
		
		\caption{Visualisation of two trajectories of prolate (a) and oblate (b) particle from experiments. Their duration is $215 \tau_{\eta}$ for $\alpha=6$ (rod) and $183\tau_{\eta}$ for $\alpha=1/6$ (disk). PDF of the square tumbling rate for rods (c) and disks (d): RB experiments (empty circles), DNS results (red line), Jeffery orbits in a steady plane shear flow (black line). For comparison experimental results in HIT from Ref.~\citep{ParsaPRL2012} at $\alpha=5, Re_{\lambda} =161$ (c) and  Ref.~\citep{Marcus2014} at $\alpha=1/10, Re_{\lambda} = 91$ (d) are shown.}
		\label{fig:1}				
	\end{center}
\end{figure}
\subsection{Experiments}\label{sec:experiments}
The experiments are carried out in a cubic Rayleigh-B\'enard cell, with side $H=24~ cm$. The dynamics of this system is determined by two control parameters, the Rayleigh number $Ra=\beta g \Delta TH^3/(\nu\kappa)$ and Prandtl number $Pr=\nu/\kappa$. Here $\beta$ is the thermal expansion coefficient, $g$ the acceleration due to gravity, $\Delta T$ the bottom-top thermal gap, $\nu$ the kinematic viscosity and $\kappa$ the thermal diffusivity. The experiments presented here have been conducted by means of a solution of glycerol with $55\%$ by weight at a mean temperature $T_m=40^oC$ which corresponds to  a Prandtl number $Pr\simeq37$. The temperature difference of the bottom and top plates in the experiments is $17.4^o C$ which corresponds to $Ra=4\times10^9$, with which the Boussinesq approximation is valid. The cell is made by acrylic foils and they are sandwiched by two thick copper top and bottom plates. The temperature of the top plate is kept constant by a refrigerated liquid circulating bath. The bottom plate is powered by a power supply to provide a constant heat flux. The  density of the fluid at $T_m$ is approximately equal to  the one of the polyamide particles seeding the flow ($\rho_p=1.13g/cm^3$) so that they can be taken on average as neutrally buoyant. 
\textcolor{black}{The particles are non-spherical and axisymmetric in shape. They are either} rod-like  with $0.45~mm$ diameter and aspect ratio $\alpha=6$, or disk-like with $3~mm$ diameter and aspect ratio $\alpha=1/6$. These shapes are produced respectively by cutting a long polyamide thread and by cutting out equal disks from a polyamide sheet. The uncertainly in length is 0.2 mm for rods which corresponds to an aspect ratio uncertainty of +/- 7\%, and it is much smaller for disks as they were cut from a polyamide sheet by means of stainless-steel tube.
 \textcolor{black}{The fluid-particle suspension is highly diluted, around 1 particle in a volume of} \textcolor{black}{$(34\eta)^3$} \textcolor{black}{ global dissipative units, corresponding to a volume fraction $O(10^{-5})$. In such condition the effect of particles on the flow is negligible.} We evaluate the global energy dissipation rate as $\overline{\epsilon}=RaPr^{-2}(Nu-1)\nu^3/H^4$ \citep{ShraimanSiggia1990}. The Nusselt number $Nu$ is measured based on the relation $Nu=QH/(\chi\Delta T)$, where $Q$ is the time-averaged total heat-flux through the system and $\chi$ \textcolor{black}{the} thermal conductivity of the fluid. The global dissipative length and time scales measure respectively, $\eta=(\nu^3/\overline{\epsilon})^{1/4} = 1.9\ mm$ and $\tau_{\eta}=(\nu/\overline{\epsilon})^{1/2} =1.1\ s$.
In such conditions the particle translational and rotational response times \textcolor{black}{estimated as in Ref.~\citep{VothARFM2017}} are sufficiently small ($\sim 10^{-2} \tau_{\eta}$) to neglect the effect of inertia on the dynamics. The three-dimensional particle tracking is performed by means of two cameras pointing to the system horizontally from orthogonal directions. 
The imaging volume covers $89\%$ of the total volume, meaning that on average the field of view reaches a distance of $4.5\ mm$ from the walls. This wide field of view allows to access both the dynamics in the kinetic boundary layers (BL), whose thickness is $\delta_{BL} = H/\sqrt{Re} \simeq 10\ mm$ (here $Re=u_{rms}H/\nu \simeq 545$) and the cell-size \textcolor{black}{large scale circulation (LSC)} flow structure. 
 \textcolor{black}{The individual three-dimensional (3D) particle trajectories are measured by combining the synchronised records from each camera using the reconstruction methods presented in \citep{MathaiExpFLu2016} for the position and in \citep{ParsaPHD2013} for the orientation. First, the particles are identified by finding the best matching in the vertical displacement and the vertical acceleration between each imaged particles in the two temporal series of frames. This allows to access the 3D position and velocity time-series for each individual particle. Second, one makes use of projective geometry to identify the actual particle orientation in space. For each identified particle one determines two planes, each plane lying on i) the line connecting the camera position to the particle position and ii) the perceived orientation of the particle symmetry axis in the frame. The intersection of such two planes allows to determine the particle orientation in the 3D. However, for measurements employing only two cameras, one shall carefully take into account the existence of an ambiguity when the particle is lying the epipolar plane, such an ambiguity is just transient and can be resolved by a polynomial fitting of the orientation in the frames before and after the undetermined event.}
The typical duration of the recorded particle trajectories is $\mathcal{O}(10)$ large eddy turnover times, $T=H/u_{rms}=26.3$ s.

\subsection{Simulations}\label{sec:simulation}
In the present study we perform two series of simulations of anisotropic particles advected by a flow: i) The first set concerns particles that are evolving in a Rayleigh-B\'enard flow with similar geometrical and physical parameters as the experiments, $Ra=10^9$, $Pr=40$; ii) The second set is for particles in a statistically homogeneous and isotropic turbulent flow at Taylor-Reynolds number $Re_\lambda=32$, $i.e.$, close to the one estimated in the center of the RB system. The need for simulation of particles in a HIT flow will be clarified further below. We adopt an Eulerian-Lagrangian model, meaning that we describe the flow evolution by means of fluid dynamics equations, while the particle trajectories (the evolution of their position and orientation) are described in a co-moving frame by ordinary differential equations. Furthermore, we assume that only the evolution of the fluid affects the dynamics of the particles, i.e., we adopt a so called one-way coupling approximation.

The flow in the RB cell is described by the Oberbeck-Boussinesq equations:
\begin{eqnarray}
\frac{\partial\textbf{u}}{\partial t}  + \textbf{u}\cdot {\nabla}   \textbf{u} &=& -   \frac{1}{\rho}{\nabla} p + \nu\ \nabla^2   \textbf{u} + \beta g (T-T_m)  \textbf{\^y}, \label{eq:RB}\\
{\nabla}  \cdot  \textbf{u} &=& 0, \label{eq:div1}\\
\frac{\partial T}{\partial t} + \textbf{u}\cdot {\nabla}   T &=&  \kappa\ \nabla^2   T, \label{eq:T}
\end{eqnarray}
where $\textbf{u}(\textbf{x},t)$ and $T(\textbf{x},t)$ are respectively the velocity and temperature fields, $\textbf{\^y}$ is the unit vector which points upwards, and the parameters are the kinematic viscosity ($\nu$), the thermal diffusivity ($\kappa$),
the reference liquid density ($\rho$) at temperature $T_m$, the thermal expansion coefficient with respect to the same temperature $(\beta)$ and finally the intensity of gravitational acceleration $(g)$. The boundary conditions for the velocity are no-slip on all cell faces. The horizontal top and bottom plates are isothermal and the lateral boundaries are adiabatic.

The HIT flow is described by the the incompressible Navier-Stokes equations driven by an external large-scale random force with constant global energy input capable to produce and sustain a statistically homogeneous and isotropic turbulent flow (same as in DNS in \citep{mathai2016microbubbles}). The boundary conditions are periodic in all directions.
\begin{figure}
	\begin{center}
		\includegraphics[width=1.0\columnwidth]{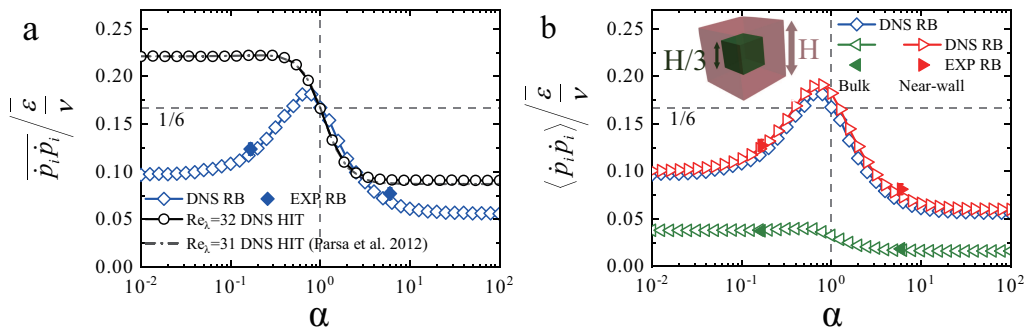}
		\caption{Mean squared tumbling rate as a function of aspect ratio $\alpha$ with global energy dissipation rate $\overline{\epsilon}$ as normalisation scale. a) Global measurements for RB DNS at $Ra=10^9$ and $Pr=40$ (blue diamond), HIT DNS at  $Re_\lambda=32$ (black circles), and for comparison HIT DNS at $Re_\lambda=31$ from Ref.\citep{ParsaPRL2012} (grey dashed-dotted line). Filled symbols are experimental measurements in RB for $\alpha=6$ and $1/6$. b) Local measurements in bulk and near-wall subdomains with the same normalisation as in panel a). \textcolor{black}{Error bars on experimental data points are determined from the absolute difference of the same measurement in two equal data subsets.}
		}
		\label{fig:2}
	\end{center}
\end{figure}

The two above described flows are seeded by a fixed number of particles. The particles are assumed to be smaller than the typical dissipative scale of the flow, so that they experience just linear flow variations across their length and width. The effects due to inertia or particle size are neglected both in their translational and rotational dynamics. In such simplified conditions, the dynamic of particles is described by the following set of equations \citep{Jeffery1922}:
\begin{eqnarray}
\dot{\textbf{r}} &=& \textbf{u}(\textbf{r}(t),t)\label{eq:part}\\
\dot{\textbf{p}} &=& \Omega \textbf{p} + \tfrac{\alpha^2 - 1}{\alpha^2+1} \left( \mathcal{S}\textbf{p} - (\textbf{p}\cdot \mathcal{S}\textbf{p})  \textbf{p} \right) \label{eq:Jeffery3d},
\end{eqnarray}
where $\textbf{r}(t)$ and $\textbf{p}(t)$ identify the particle position and orientation respectively, 
$\alpha$ is the aspect ratio of the particles, $\mathcal{S} = ({\nabla}   \textbf{u}+ {\nabla}   \textbf{u}^T)/2$ and $ \Omega=({\nabla}   \textbf{u}-{\nabla}   \textbf{u}^T)/2$ represent the symmetric and anti-symmetric components of the fluid velocity gradient tensors at the particle position, ${\nabla}   \textbf{u}$.
\textcolor{black}{In our experiments the size of particles is smaller or comparable to the dissipation length scale of the flow, supporting the assumption that Jeffery's equation is an appropriate model to describe the angular dynamics of the particles in the present conditions, see also \citep{Ravnik2018} for a more detailed discussion.} \textcolor{black}{For the present simulations a resolution of $512^3$ equally spaced grid-points is used for the RB flow at $Ra = 10^9$, $Pr = 40$, while a $256^3$ grid is used for the HIT flow at $Re_\lambda = 32$.}
In all simulations the anisotropic particles cover a wide range of aspect ratios from thin disks to slender fibers, $\alpha \in \left[10^{-2},10^2\right]$. 
The above described models are simulated numerically by means of a tested code \citep{Calzavarini_SI2019} which makes use of a Lattice Boltzmann Method (LBM) algorithm for the treatment of the fluid and thermal part and of a second-order in time tracking algorithm with trilinear interpolation of fields at the particle positions.  The code has been used for the simulation of convection in a closed cavity \citep{JiangPRE2019}, convection with particles \citep{calzavarini2020anisotropic} and HIT flows with inertial particles \citep{mathai2016microbubbles,Calzavarini2018microprobes}.

\section{Results and discussion}\label{sec:results}

A visualisation of two typical trajectories of prolate and oblate particles from the experiments is shown in Figure~\ref{fig:1}(a-b). 
The evolution of the particle center of mass gives the indication of the large scale circulation which occurs in the RB cell in the present conditions. 
The particle orientation, also visually rendered in Figure~\ref{fig:1}(a-b), identified by the unit vector $\textbf{p}(t)$,  evolves smoothly in the bulk and changes abruptly mostly in correspondence of the top and bottom regions (although particle-wall collisions do not occur). 
A quantitative measurement of these changes is obtained via the probability density function (PDF) for the instantaneous normalized quadratic tumbling rate, $\dot{p}_i \dot{p}_i/\overline{\dot{p}_i \dot{p}_i}$. The PDFs, Figure~\ref{fig:1}(c-d), show exceptionally large fluctuations, up 60 times the mean quadratic tumbling value. For comparison we display on the same figure the expected PDF for a set of initially random uniformly oriented particles of the corresponding aspect ratios tumbling in a stationary plane shear flow, \textit{i.e.}, performing so called Jeffery orbits. Note that the latter PDF has a finite support due to the periodic evolution of the particle orientation, and it has a much shorter tail. Additionally, we remark that the shape of these curves: i) does not seems to carry a particle aspect ratio dependence ii) agrees well with previous experimental measurements \citep{ParsaPRL2012,Marcus2014} performed in developed turbulent flows at much higher Reynolds numbers iii) are in excellent agreement with the DNS. The latter point give us confidence in relying on numerical results for further analysis. 

We now look at the dependence of the mean quadratic tumbling as a function of the aspect ratio; \textcolor{black}{the symbol $\overline{ \vphantom{M}\ldots}$ denoting in the following global volume and time averages}. Figure \ref{fig:2}(a) presents the measurements from numerics in the aspect ratio range $\alpha \in \left[10^{-2},10^2\right]$ together with the ones from the experiments for oblate and prolate particles, which again nicely agree with each other. More importantly the behaviour is compared with the measurements of the same quantity in HIT, here DNS at $Re_{\lambda}=32$ and results from \citep{ParsaPRL2012} in the same conditions. 
The tumbling rate in RB is for all the aspect ratios slower than the one observed in the HIT flow, pointing to the existence of either a stronger preferential alignment with the underlying flow or to an intrinsic smaller variability of the velocity gradient fluctuations.

A major difference between RB turbulence and HIT concerns  the non-homogeneity caused by the presence of geometrical boundaries. 
Therefore, \textcolor{black}{following the decomposition approach in \citep{grossmann_lohse_2000}} we divide the volume in two subsets, a central cubic volume of side $H/3$ denoted as ``bulk'' and the complement of it denoted as ``near-wall'' region (see cartoon in Figure~\ref{fig:2}(b)). 
Figure ~\ref{fig:2}(b) shows the computed mean tumbling rates in the two regions both form experiments and simulations \textcolor{black}{(note that we use the the symbol $\langle \ldots \rangle$ to indicate a local in space average)}. Two observations are in order. First, this analysis confirms the strong inhomogeneity of the RB flow. The global tumbling rate appears to be dominated by the near-wall region, where it is twice as strong as in the bulk. Second, we observe very different trends for the quadratic tumbling rate versus $\alpha$ in the two regions. 
\begin{figure}
	\begin{center}
		\includegraphics[width=0.55\columnwidth]{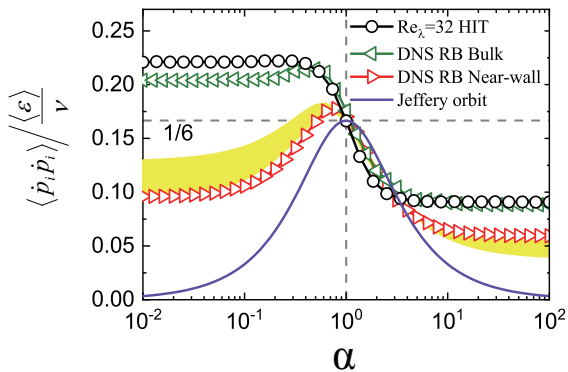}
		\caption{Local measurements of mean squared tumbling rate in  bulk and near-wall regions with the respective local energy dissipation rates normalizations ($\langle\epsilon\rangle$ standing here either for $\langle\epsilon\rangle_{\textrm{bulk}}$ or $\langle\epsilon\rangle_{\textrm{near-wall}}$) as a function of aspect ratio $\alpha$. For comparison we show HIT DNS at $Re_\lambda=32$ (black circles), Jeffery orbits (violet line) and a best fit of the linear-combination of the two latter curves:  \textcolor{black}{$a\times [ \langle \dot{p}_i \dot{p}_i \rangle \nu / \langle \epsilon \rangle]_{\textrm{HIT}} + (1-a)\times [\langle \dot{p}_i \dot{p}_i \rangle \nu / \langle \epsilon \rangle]_{\textrm{Jeffery}}$} with coefficient $a=0.5 \pm 0.08$.
		}
		\label{fig:2-2}
	\end{center}
\end{figure}
In order to make sense of these two behaviours we normalise the mean tumbling rate with respect to the local dissipation scales in the two subregions, \textit{i.e.} the energy dissipation rate $\langle \epsilon \rangle$ is here computed either in bulk, or near-wall domains, see Figure~\ref{fig:2-2}.  For the bulk we observe a remarkable match with the measurements in HIT, proving that 
the flow gradient properties in the core of the RB system, at this Ra number, are the same as in nearly homogeneous and isotropic turbulent flows.
Interestingly, when the local energy dissipation rate rescaling is adopted the near-wall tumbling rate falls below the bulk one. This behaviour can be contrasted with the one of an initially isotropic set of particles evolving in a steady plane shear flow, denoted as Jeffery orbits in the same figure \footnote{Note that the behaviour for Jeffery orbits in Figure~\ref {fig:2-2} corrects the one provided in Ref. \citep {ParsaPRL2012} Figure~4, which was affected by a normalisation error, hence the mentioned curve in that work should be multiplied by a factor 4/3. The mean normalized tumbling rate for Jeffery orbits is well approximated by $\alpha /(3\alpha ^2+3)$.}.

It is now tempting to speculate that the behaviour observed in the near-wall (and indeed globally dominant in the RB system) is a superposition of turbulence and steady shear.  A test of this guess, based on a simplistic linear combination of the two \textcolor{black}{mentioned} tumbling rate behaviours does not fall far from the actual shape (see shaded region in Figure \ref{fig:2-2}). The linear combination fit provides an estimate for the background shear intensity comparable to the one of turbulent fluctuations.
More importantly, it accounts for the observed local maximum, that occurs for aspect ratios just below the unit value. 	
\begin{figure}
	\begin{center}			
		\includegraphics[width=1\columnwidth]{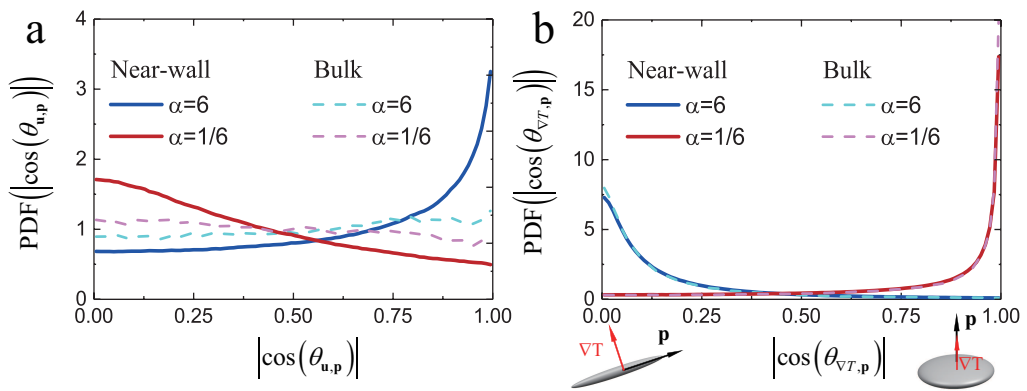}
		\caption{(a) PDF of the absolute value of the cosine of the angle of particle orientation with the fluid velocity, $|\cos \theta_{\textbf{u},\textbf{p}}| =  |\textbf{p} \cdot \textbf{u}|/||\textbf{u}||$; (b) the same for the temperature gradient, $|\cos \theta_{{ \nabla}T,\textbf{p}}| =  |\textbf{p} \cdot  \nabla T|/||{\nabla}T||$. 
		}
		\label{alignments}			
	\end{center}
\end{figure}

In order to support the hypothesis that a background shear affects the observed global tumbling rate we study particle orientation with respect to the flow velocity. Figure \ref{alignments}(a) shows by means of DNS data that $\alpha=6$ rods align with the velocity in the near-wall region, while $\alpha=1/6$ disks are orthogonal to it; on the contrary in  the bulk any preferential alignment vanishes. Note that the PDF of the modulus of the cosine is considered here for symmetry reason: due to the \textcolor{black}{fore-aft} symmetry of particles the parallel or antiparallel orientation with respect to a vector are equivalent. This qualitatively agrees with Jeffery orbits phenomenology in a shear flow \citep{Jeffery1922}. \textcolor{black}{Additionally, we also verified that rods/disks are biased towards parallel/perpendicular orientations to the cell walls,  in agreement with the near-wall dynamics in turbulent channel flows \citep{MarchioliPF2010,Challabotla2015}, see} 
\textcolor{black}{ \ref{appA} for details}.

In Figure~\ref{alignments}(b) we also show that disk-like particles gets strongly aligned with the temperature gradient, $\bm{\nabla} T$, while rod-like particles stay \textcolor{black}{weakly but preferentially} orthogonal to it. 
This feature occurs equally well in the bulk and in near-wall and it is related to a similarity between the orientation equation for the particle and the evolution equation for the gradient of a scalar field advected by the fluid, as first proposed in \citep{calzavarini2020anisotropic}.
\textcolor{black}{Indeed, eq. (\ref{eq:Jeffery3d}) can be rewritten in terms of the evolution of an auxiliary non-unit vector, $\textbf{q}(t)$,\citep{Szeri1993}: 
\begin{equation}\label{eq:szeri}
\textbf{p}(t)=\frac{\textbf{q}(t)}{||\textbf{q}(t)||}, \qquad \dot{\textbf{q}}(t) = \left( \bm{\Omega} +  \frac{\alpha^2-1}{\alpha^2+1} \bm{\mathcal{S}} \right)\textbf{q}.
\end{equation}
The limit of a thin disk, $\alpha \to 0$, in (\ref{eq:szeri}) leads to an equation that apart from the diffusive term is formally identical to the one of the gradient of a scalar field, $T$, following the advection diffusion equation:
\begin{eqnarray}
\qquad \dot{\textbf{q}}_{\alpha=0}(t) = - \nabla\textbf{u}^T\ \textbf{q}_{\alpha=0},\qquad
\dot{\bm{\nabla}T} = - \nabla\textbf{u}^T\ \bm{\nabla}T + \kappa \Delta \bm{\nabla}T \label{DT}
\end{eqnarray}
Similarly the opposite limit of a rod, $\alpha \to \infty$, as first pointed out by \citep{PumirWilkinson2011},  shares an analogous similarity with the vorticity ($\bm{\omega}$)  equation of motion: 
\begin{eqnarray}
\qquad \dot{\textbf{q}}_{\alpha=\infty}(t) =  \nabla\textbf{u}\ \textbf{q}_{\alpha=\infty},\ \ \ \qquad
\dot{\bm{\omega}} = \nabla\textbf{u}\ \bm{\omega} + \nu \Delta \bm{\omega}.
\end{eqnarray}
Note that while the former equivalence becomes exact in the limit of $Pr \to \infty$, the latter one occurs in the opposite limit $Pr \to 0$. As a consequence one shall expect a correlation between the rotational dynamics of the temperature gradient with the orientation of a disk; such a correlation is only approximate owing to the presence of thermal dissipation and to the different initial conditions for the particle and thermal gradient orientations. However, the correlation effect is particularly evident in the present flow due to the high value of $Pr$ which makes the decorrelating effect of thermal diffusivity negligible.  We also observe that $\textbf{q}_{\alpha=\infty}\cdot \textbf{q}_{\alpha=0} = const.$, suggesting that rod like particles would be most likely not in line, with the thermal gradient direction. Appendix \ref{appB} reports complementary measurements on particle preferential alignment with the vorticity and rate-of-strain eigenvalues both in RB and HIT.
}\\

\begin{figure}
	\begin{center}
		\includegraphics[width=0.6\columnwidth]{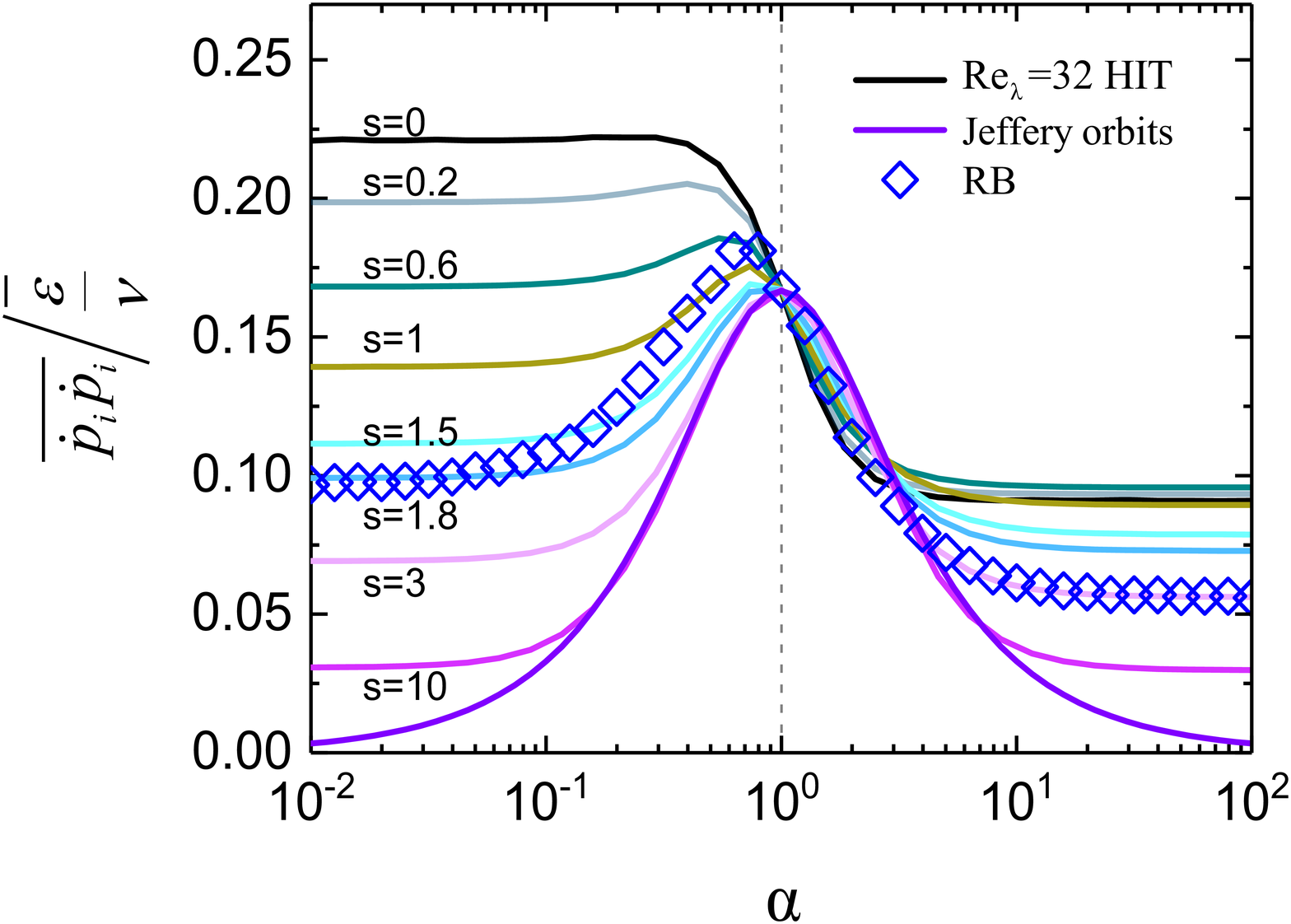}
		\vspace{-0.5cm}
		\caption{Mean quadratic tumbling rate versus the particle aspect ratio as obtained evolving a particle ensemble with (\ref{eq:synt}) where the intensity of the background shear field is $s$ times the average intensity of the  HIT velocity gradient fluctuations. The pure HIT case $s=0$ (black line), and the pure shear case, Jeffery orbits $s=\infty$ (violet line) and the measurements in RB from DNS (diamonds) are also drawn.
		}
		\label{fig:5}			
	\end{center}
\end{figure}

The measurements presented so far support the hypothesis that the global tumbling rate in RB is affected by a persistent background shear flow coexisting with turbulent fluctuations.
In order to better study the effect of this superposition and the sensitivity of particle tumbling  to the relative intensity of shear as compared to turbulence, we build a synthetic velocity gradient field as follows:
\begin{equation}
\label{eq:synt}
\nabla \textbf{u}_{s}(t) =  \frac{1}{\sqrt{1+s^2}}\left(  \nabla \textbf{u}(t) + 
\left( \begin{array}{ccc} 
0 & s\  \tau_{\eta}^{-1} & 0 \\ 
0 & 0 & 0\\
0 & 0 & 0
\end{array}\right) \right).
\end{equation}
Here $\nabla \textbf{u}(t)$ comes from a DNS record of the evolution of the gradient tensor along Lagrangian trajectories in HIT, while $s$ is an adjustable shear intensity coefficient in $\tau_{\eta}$ units. Note that the overall normalisation is constructed in such a way that the mean dissipative time ($\tau_{\eta}$) of the resulting gradient field $\nabla \textbf{u}_{s}$ is the same as the one of the original HIT field $\nabla \textbf{u}(t)$. This allows to compare the shear intensity with the amplitude of the global field. As a caveat, we note that (\ref{eq:synt}) is a rough approximation of a realistic turbulent shear flow, where velocity gradient fluctuations have specific signatures \citep{Pumir2017}, and obviously also as a simplistic approximation of a RB flow.  
The mean tumbling rate of an initially random uniformly oriented set of particles evolved along many of such gradient trajectories is reported in Figure~\ref{fig:5}. The 
family of curves, corresponding to different values of $s$, show similar features with the ones of the quadratic tumbling rate measured in the RB cell. Interestingly, this model provides an estimate, $s^* = 2 \pm 0.3$, of the background shear-rate intensity as compared to the turbulent velocity gradient fluctuations.  The estimated associated time scale ($\tau_{s}$ for shear) expressed in the dissipative-time of the synthetic global field ($\tau_{\eta}$) turns out to be  $\tau_{s} = \tau_{\eta} \sqrt{1+s^{*2}}/s^*  \simeq 1.1 \pm 0.1 \tau_{\eta}$.  
We note that this intensity is incompatible with a persistent shear-rate determined by the LSC  turnover time, because here $T=H/u_{rms} \simeq 24 \tau_{\eta}$, while it is much closer to the time-scale associated to the kinetic boundary layers  $\tau_{BL} = \delta/u_{rms} = T Re^{-1/2} \simeq 1.6 \tau_{\eta}$. 

Finally, it is tempting to ask what will be the fate of the mean quadratic tumbling intensity \textcolor{black}{as a function of} $\alpha$ at varying the Rayleigh number. This can be guessed dimensionally by considering (see \citep{ShraimanSiggia1990,grossmann_lohse_2000} for the estimates) that $\langle \epsilon \rangle_{BL} \sim \nu (u_{rms}/\delta)^2$ and occupies a volume $V_{BL}=(H^3 -(H-2\delta)^3)$ while  $\langle \epsilon \rangle_{bulk}\sim u_{rms}^3/(H-2\delta)$ occurs in a volume $V_{bulk}=(H-\delta)^3$ (where bulk denotes here the total volume minus the one occupied by BLs). With this we evaluate $s^* \sim \sqrt{\langle \epsilon \rangle_{BL} V_{BL}}/\sqrt{\langle \epsilon \rangle_{bulk} V_{bulk}} \sim Re^{-1/2}$, which is bound to decrease to zero as $Ra$ increases (because $Re\sim Ra^{1/2}$). Similarly, $(\tau_{s}/\tau_{\eta})^2  \simeq (\overline{\epsilon} H^3)/(\langle \epsilon \rangle_{BL} V_{BL})$ increases with $Ra$, meaning that the relative importance of the \textcolor{black}{mean shear} will reduce with respect to turbulence. \textcolor{black}{However, we remark that the above dimensional argument is speculative and shall be put under scrutiny by future experimental/numerical studies capable to access higher $Ra$ number conditions.}\\

\section{Summary}\label{sec:summary}
In summary, the orientation dynamics of inertialess anisotropic particles advected by Rayleigh-B\'enard turbulent convection is studied by experiments and simulations. 
The heavy tail distributions of tumbling rate show that the flow possesses similar extreme small-scale fluctuations as in HIT, despite the difference in the Reynolds number. 
On the contrary,  the mean tumbling rate dependence on the particle aspect-ratio displays a maximum for weakly oblate particles ($\alpha \lesssim 1$), rather than the universal monotonic decreasing trend observed in homogeneous and isotropic turbulence. The mean tumbling rate is highly inhomogeneous across the system, larger in the outer regions, weaker in the bulk where remarkably the mean tumbling rate behaviour of HIT is fully recovered (both in magnitude and $\alpha$ dependence). We show that the peculiar trend observed in the RB system is due to the combined effect of turbulent fluctuations and a persistent mean shear flow component. This is supported by the observed parallel (orthogonal) alignment of prolate (oblate) particles with the local fluid velocity, a trait of Jeffery orbits in steady plane shear flows.
By means of a synthetic shear turbulence model we estimate that the mean shear-rate intensity necessary to produce the observed tumbling dynamics in the RB cell in the present conditions is of the order of the inverse global Kolmogorov time scale.

We revealed that the rotation statistics of particles advected by a realistic wall-bounded thermally-driven turbulent flow is on a global level fundamentally different from the one measured in unbounded turbulent flows, due to the effect of  persistent torques induced by the wall boundary layers.
These results have implications for the study of particle rotational dynamics in the wider class of turbulent flows with shear, where they can improve the interpretations of already available measurements.  Future studies should focus on a more detailed quantification of rotation rates of anisotropic particles as a function of the distance from the wall in turbulent flow conditions. Such a characterization demands for extremely long simulations/experiments in order to have converged statistics. Finally, the revealed preferential alignments with velocity and temperature gradient may be relevant for the modelling of the dynamics of aspherical organism transported by shear dominated turbulence \citep{GuastoARFM2012}, with or without thermal effects \citep{BorgninoPRL2019}.

\section*{Acknowledgements}
We gratefully acknowledge V.~Mathai and X.~Ma for their help with experiments, and Detlef Lohse for useful discussions. EC acknowledges useful discussions with B. Mehlig and G. Verhille.
This work is financially supported by the Natural Science Foundation of China under Grant No.~11988102, 91852202, 11861131005 and 11672156, and Tsinghua University Initiative Scientific Research Program (20193080058).

\appendix
\section{Additional measurements on particle alignment}
\subsection{Alignment with RB cell walls}\label{appA}
\begin{figure}
	\begin{center}			
		\includegraphics[width=0.8\columnwidth]{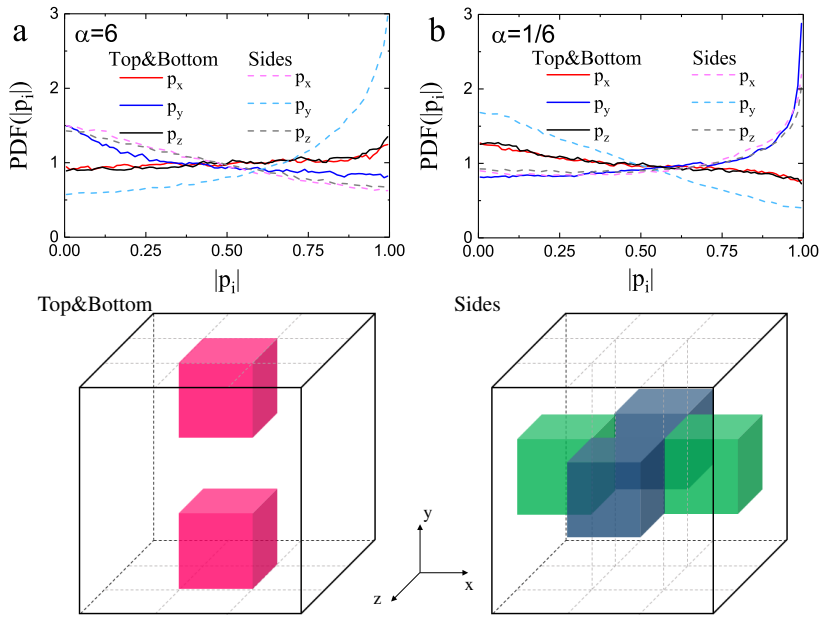}
		\caption{Decomposition of the near-wall domain (as defined in the main paper) in sub-regions localised either next to the horizontal or vertical boundaries. We denote them as ``Top \& bottom" or ``Sides" domains  (see cartoon on bottom of the figure). Alignment of particles with wall directions for  ``Top \& bottom" wall (solid lines) and the  ``Sides" wall (dashed lines) for (a) $\alpha = 6$ and (b) $\alpha = 1/6$. Note that the vertical direction in our system is along the $y$ axis.}
		\label{wall-alignments}			
	\end{center}
\end{figure}

%Appendix
In order to further support the hypothesis that a background shear flow affects the global tumbling rate in the RB cell we study the alignments of particle orientation with the wall directions. 
For this analysis \textcolor{black}{we have further decomposed} the near-wall domain  in sub-regions localised either next to the horizontal or vertical boundaries (see cartoon on Fig. \ref{wall-alignments}). We denote them as ``Top \& bottom" or ``Sides" domains.
Our DNS measurements provide evidence that prolate particles (here $\alpha=6$) are preferentially parallel to walls while oblate ones ($\alpha=1/6$) are mostly orthogonal to them, see Fig. \ref{wall-alignments}. These observations are consistent with the preferential orientations detected close-to-wall in simulations of anisotropic particles in  turbulent channel flows, see \citep{MarchioliPF2010} for rods and \citep{Challabotla2015} for disks.

\subsection{Alignment with velocity, vorticity, rate-of-strain and thermal gradient}\label{appB}
\begin{figure}
	\begin{center}
		\includegraphics[width=1.0\columnwidth]{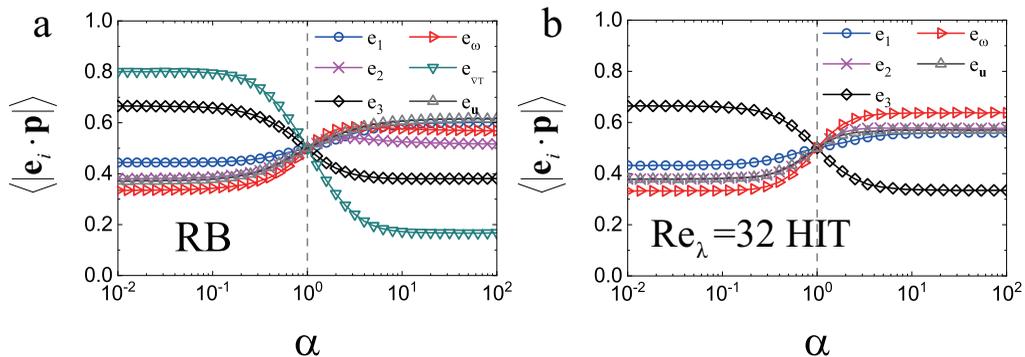}
		\caption{a). Alignments of particles with unit velocity vector $\textbf{e}_u$, vorticity vector $\textbf{e}_\omega$, rate-of-strain eigenvectors $\textbf{e}_{1,2,3}$ and thermal gradient $\textbf{e}_{\nabla T}$ in RB system as a function of aspect ratio $\alpha$. b). Alignments of particles with unit velocity vector $\textbf{e}_u$, vorticity vector $\textbf{e}_\omega$, rate-of-strain eigenvectors $\textbf{e}_{1,2,3}$ in HIT ($\rm Re_\lambda=32$) as a function of aspect ratio $\alpha$.}
		\label{cosinus}			
	\end{center}
\end{figure}

We look here at the mean alignment, more precisely at the absolute value of the cosine angle between the particle director $\textbf{p}$ and several physical quantities: the flow velocity $\textbf{u}$ , the vorticity $\bm{\omega}$, the eigenvectors of the rate-of-strain tensor $\textbf{e}_1,\textbf{e}_2,\textbf{e}_3$ and the temperature gradient $\bm{\nabla}T$.
Figure \ref{cosinus}~(a) shows the measurements for the RB system from DNS at $Ra=10^9$, $Pr=40$, while Fig.\ref{cosinus}~(b) shows the corresponding data for the HIT flow at $Re_{\lambda}=32$, i.e. at comparable Reynolds number as the one estimated for the bulk. In the RB system the \textcolor{black}{strongest alignment} is observed for oblate particles with the temperature gradient direction. Prolate particles maximally align with the velocity field. It appears that they are also correlated with $\textbf{e}_1$ and to a lesser extent to the vorticity. This is contrasted by the measurements in isothermal HIT, where prolate particles maximally align with the vorticity orientation and to a lesser extent with $\textbf{e}_2$ as first proposed in \citep{PumirWilkinson2011} and measured in \citep{ni_ouellette_voth_2014,ni_kramel_ouellette_voth_2015,Byron2015}. We note however that also the alignment with the velocity is not negligible. This is to be attributed to the small turbulence level of the present conditions; ee have verified that at higher $Re$ such a correlation vanishes.

%\bibliographystyle{jfm}
%\bibliography{rotation_jfm}

\end{document}